\documentclass[useAMS,usenatbib]{mn2e}
\usepackage{graphicx}
\usepackage{lscape}
\usepackage{times}
\usepackage{amsmath}
\usepackage{amsfonts}
\usepackage{amssymb}
\usepackage{multirow}
\usepackage{longtable}
\usepackage{afterpage}
\RequirePackage{color}

\title{On the origin of W UMa type contact binaries - a new method for computation of initial masses}
\author[M. Y\i ld\i z and T. Do\u{g}an]{M. Y\i ld\i z$^{}$\thanks{E-mail:
mutlu.yildiz@ege.edu.tr} and T. Do\u{g}an$^{}$\\
$^{}$Department of Astronomy and Space Sciences, Science Faculty, Ege University, 35100, Bornova, \.Izmir, Turkey.}

\begin{document}
%\onecolumn
\date{}

\pagerange{\pageref{firstpage}--\pageref{lastpage}} \pubyear{2012}

\maketitle

\label{firstpage}

\begin{abstract}
W UMa type binaries have two defining characteristics. These are (i) the
effective temperatures of both components are very similar, and (ii) the secondary 
(currently less massive) component is overluminous for its current mass. 
We consider the latter to be an indication of its
mass before the mass-transfer event. For these stars, we define a mass difference
($\delta M$) between the mass determined from its luminosity and the present mass determined
from fitting the binary orbit. We compare the observed values of the mass difference
to stellar models with mass-loss. The range of initial
secondary masses that we find for observed W UMa type binaries is $1.3-2.6 \mathrm{M}_{\odot}$. We discover
that the A- and the W-subtype contact binaries have different ranges of
initial secondary masses. Binary
systems with an initial mass higher than $1.8 \pm 0.1\mathrm{M}_{\odot}$ become A-subtype while systems
with initial masses lower than this become W-subtype. Only 6 per cent of systems
violate this behavior. We also obtain the initial masses of the primaries using the
following constraint for the reciprocal of the initial mass ratio: $0 < 1/q_i < 1$. The range of
initial masses we find for the primaries is $0.2-1.5 \mathrm{M}_{\odot}$, except for two systems.
Finally in comparing our models to observed systems, we find evidence that the mass
transfer process is not conservative. We find that only 34 per cent of the mass from
the secondary is transferred to the primary. The remainder is lost from the system.
\end{abstract}


\section{Introduction}
The evolution and structure of a star are primarily determined by its
initial mass. Therefore, the structure of stars in contact binaries (CBs) can be very
different from single stars because of mass transfer. The component stars of CBs
evolve as single stars with initial mass ($M_{\rm i}$) until 
the binary interaction occurs.
When the
mass transfer starts, the initially more massive component
($M_{\rm 2i}$, the current
secondary) loses mass that is transferred to the initially less massive component
($M_{\rm 1i}$, the current primary) and the rest leaves the system. Because of the
interaction, the mass ratio of the system reverses. After mass transfer stops or
sufficiently slows down, the secondary stars are found to be overluminous for a
main-sequence (MS) star of the same mass. The reason for this luminosity excess is that
the luminosity of such stars depends on their initial mass as much as their present
mass. In this study, we develop a method to find initial masses of the secondary
stars of W UMa type binaries in terms of the luminosity excess.

W UMa type CBs are among the best studied stars in the literature.
For about 100 CBs (51 A-subtypes and 49 W-subtypes), accurate dimensions of component 
stars from photometric and spectroscopic data are successfully derived by different investigators.
The subtypes are defined by their
light-curves (Binnendijk 1970).
We compute initial masses of secondary and primary components of these systems.
Our sample includes well-determined parameters of CBs from the compilations of 
\citet{b18}, \citet{bb1,bb2} and \citet{b21, b20}. 
%The list of CBs we used here is based on mainly the sample of CBs given by \citet{b18}.\\
%We exclude the systems for which third body is reported in the literature. 

Beside the luminosity excess of the secondary components in the W UMa type binaries,
another very important feature of these systems is that the effective temperatures, $T_{\rm e}$, of 
components are almost identical. 
However, 
the slight difference between
effective temperatures of components is systematic between the two subtypes. For the
A-subtype systems, the effective temperature of the massive component ($T_{\rm 1e}$)  is higher
than the effective temperature of the lower mass component ($T_{\rm 2e}$); 
the opposite is true for the W-subtype systems.
There is debate in the literature concerning whether the subtypes
represent an evolutionary sequence
\citep{awada,bbgaz2,eker}.  
According to \citet{b12}, however, there is no evolutionary link between 
most of the A-subtype systems and the W-subtype CBs. 
In this paper, we
aim to confirm that the subtypes are determined by their initial parameters.
If initial parameters of A- and W-subtypes are completely different, then they are not
an evolutionary sequence.

There are two evolutionary pathways that lead to a star filling its Roche lobe. 
These are nuclear and angular momentum evolution \citep{t15}. It is
thought that angular momentum evolution via magnetic braking
\citep{schatz,mestel}  is the primary mechanism for the formation and evolution of W UMa type CBs
\citep{okamot,vantve,vilhu}. 
However, both these two
mechanisms affect the kind of W UMa binaries that are formed. Also because nuclear and
angular momentum evolution are very strongly dependent
on stellar mass, one may expect that the initial masses of components of W UMa type
binaries are the key parameters for our understanding of the formation and evolution
of W UMa type binaries.

Each component of a CB fills its 
Roche lobe and is under the gravitational interaction of the other. Apart from this, the more massive 
primary component is an MS 
star and the luminosity of the secondary component is  higher than the expected MS
luminosity \citep{b19, b17, b10}. This excess may be due to large
amount of energy transfer that occurs (see Section 2) from the primary to the secondary \citep{b11, c11, b17, b10}.
 However, the evolutionary status of
CBs is still unclear.
According to \citet{b3}, the primary components of the W-subtype CBs are not evolved MS stars, 
and the A-subtype CBs are close to the terminal-age of the main sequence (TAMS). 

Although there are many studies in the literature about contact systems, the contact models are 
still not able to satisfy all the observational constraints. It is still an open question whether CBs are in thermal 
equilibrium or not. It is generally accepted that a contact configuration in thermal equilibrium is not possible and therefore CBs suffer from 
thermal relaxation oscillation \citep{c1, b14, b5, b6, b7, b8, b17,  b9, b18, b13}. However, according to some authors, 
a contact 
configuration can be in thermal equilibrium. They argue that contact systems can achieve thermal equilibrium if the secondary, 
currently less massive component, is more evolved than the massive component \citep{s1, s2, s3, s4}. In such a model, 
CB stars 
originate from detached close binaries with initial orbital periods close to a few days and the system was exposed to mass exchange 
with a mass ratio reversal in the past. According to these assumptions, energy exchange 
does not influence stellar radii and occurs 
in thermal equilibrium \citep{b7, s4}. 
Large-scale energy transfer from the primary to the secondary components solves several outstanding problems in CBs \citep{b11}. However, the problem 
of energy transfer requires an intrinsic driving mechanism and its effect on the evolution of the component stars 
and the system is unknown. 
These processes are very poorly understood because of the complexity of detailed hydrodynamic treatment of the flow and also due to the unknown 
driving mechanism. 

{ 
This paper is organized as follows. We outline the physical properties of CBs in Section 2 and
we discuss the basic properties of CBs in 
the mass-luminosity ($M-L$) and the mass-radius ($M-R$) diagrams. In Section 3, 
%we describe how
%initial masses of the secondary components may be related to their current mass and luminosity.
%In Section 4, 
we develop and apply
a comprehensive method for computation of initial masses of CBs based on stellar modelling with mass-loss.
In Section 4}, we apply this method to the CBs and discuss the results. 
Finally, in Section 5, we draw our conclusions.

\section{The physical properties of contact binaries and the problem of energy transfer}
\begin{figure*}
\includegraphics[bb=-25 340 590
480,width=140mm,angle=270]{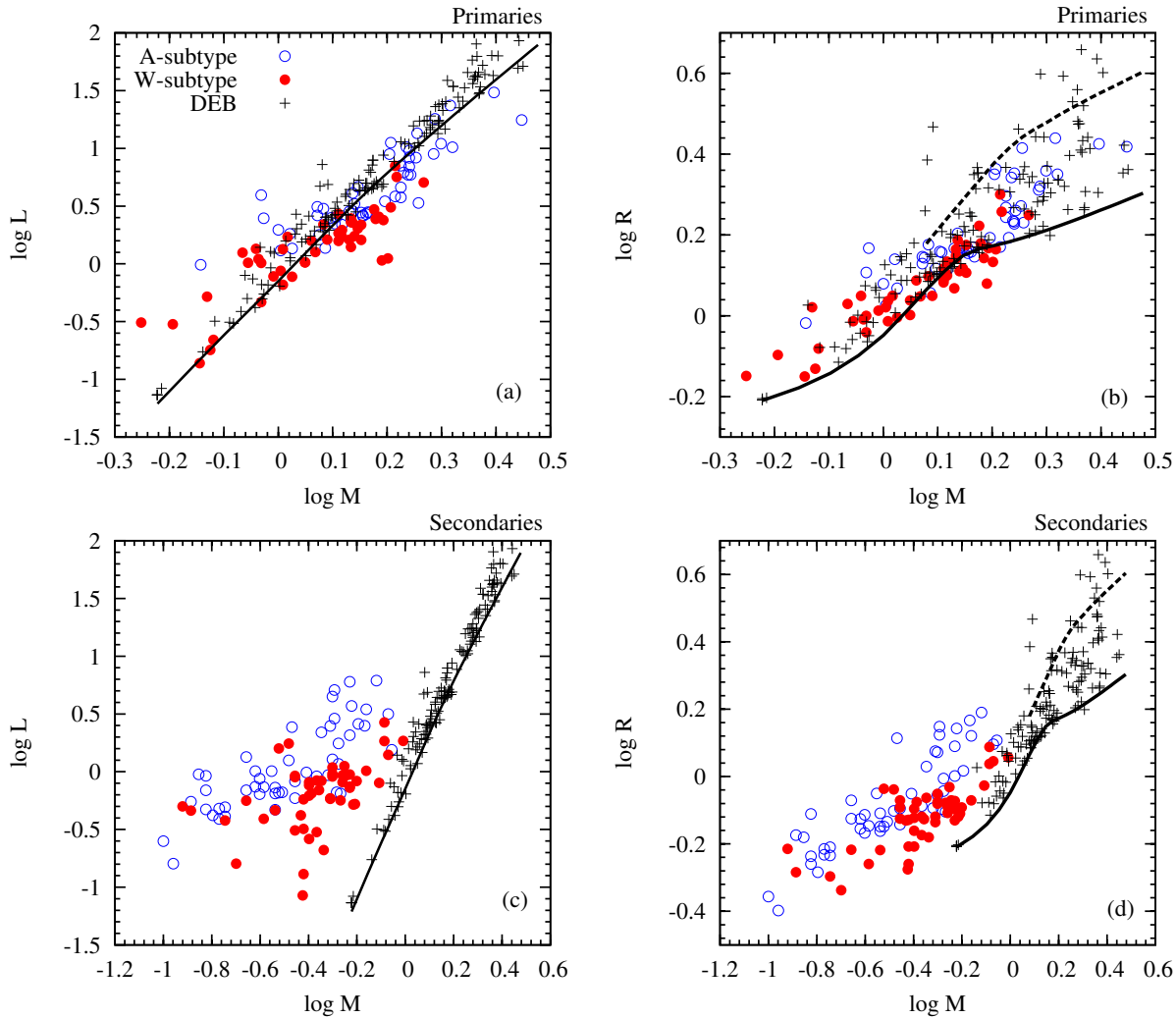}
%480,width=140mm,angle=270]{the_four_diagrams16.ps}
\caption{Mass-luminosity and mass-radius relation for the primaries of the A- (circle)
and the W- subtype (filled circle) CBs are plotted in panels (a) and (b), respectively. 
For comparison, the components of
the well-known DEBs (+) are also plotted. Solar
units are used. 
A solid line indicates the ZAMS line and a dashed line the TAMS line, obtained by using
the MESA stellar evolution code.
Panels (c) and (d) are same as (a) and (b), respectively, but for secondary components of W UMa type CBs.
}
\end{figure*}

The most reliable accurate parameters (mass, radius and luminosity) of binary stars 
are computed by using both photometric and spectroscopic data.
We compiled 100 such well known CBs from the literature.
Their basic properties are listed in Table A1 of the Appendix. 
These properties give clues to evolutionary phases of the primary components, at least.
The primary components of CBs in $M-L$ and $M-R$ diagrams are much closer to that of the 
normal MS stars than the secondary components. 
Their luminosities and radii %(circle for A-subtype and filled circle for W-subtype) 
are plotted with respect to their masses 
in Figs 1(a)  and 1(b), respectively. For comparison, the components of the well-determined 
detached eclipsing binaries (+; hereafter DEBs) are also 
included \citep{torres}. The solid line in these figures shows zero-age MS (ZAMS), 
derived from models constructed with the MESA code \citep{bb13}.
There is an agreement between the luminosities of the primaries of CBs and that of the components of DEBs. 
If we compare the radii of the primary stars of CBs in Fig. 1(b), 
we can confirm that these two groups of stars are in very good agreement with the DEB stars. 
Fig. 1(b) also shows the TAMS line
(as a dashed line). Most of the primaries of CBs occur between the ZAMS and the TAMS lines.

The  $M-L$ and $M-R$ diagrams of the secondary components of
W UMa type CBs are plotted in Figs 1(c) and 1(d), respectively.
Unlike the primaries shown in Figs 1(a) and Figs 1(b), there is no agreement between  the secondary components 
and the components of DEBs in $M-L$ and $M-R$ diagrams.
Therefore, the secondary components of W UMa type CBs must have a very different 
structure or evolutionary path.
All the secondary components are above 
the ZAMS line in both Figs 1(c) and 1(d). 
The mean masses of the secondary components of the A- and W-subtype CBs are $0.38$ and $0.47 \mathrm{M}_\odot$, respectively.
Despite these low masses, the secondary stars are very bright and very large in comparison to their
MS counterparts of the same mass. This excess in luminosity and radius may be explicable by the early evolution of 
the systems. 

In the literature, the close $T_{\rm e}$s of components of  CBs have been explained as a result of 
an energy transfer 
from the primary to the secondary star \citep{b15}.
According to this approach, the luminosity excess of the secondary components ($\delta L_2=L_2-M_2^4$) 
must compensate the lack of luminosity of the primary components ($\delta L_1=L_1-M_1^4$).
If this is the case, there must be a correlation between $\delta L_1$ and $\delta L_2$.
In Fig. 2, $\delta L_2$ is plotted with respect to $\delta L_1$. It should be noted that
$\delta L_2$ is always greater than zero and $\delta L_1$ is in general (four fifths of systems) % 19-22/98
less than zero. 
If the luminosity excess of the secondary stars 
compensates the lack of luminosity of the primary stars, the position of the CB components 
should be on the $\delta L_2=- \delta L_1$
line (solid line). There is a large scatter for the A-subtype CBs (circles) and therefore no 
evidence for the  energy transfer process is shown for this subtype. 
For the W-subtype CBs (filled circles), however, some of the systems with small values of $\delta L_1$ and 
$\delta L_2$ form a sequence on the line $\delta L_2=-\delta L_1$. 
The sequence is very clear from the origin to the point with $\delta L_2=1 \mathrm{L}_\odot$ and 
$\delta L_1=-1 \mathrm{L}_\odot$. 
The details of this range 
are also shown in Fig. 2.
\begin{figure}
\includegraphics[width=67mm,angle=270]{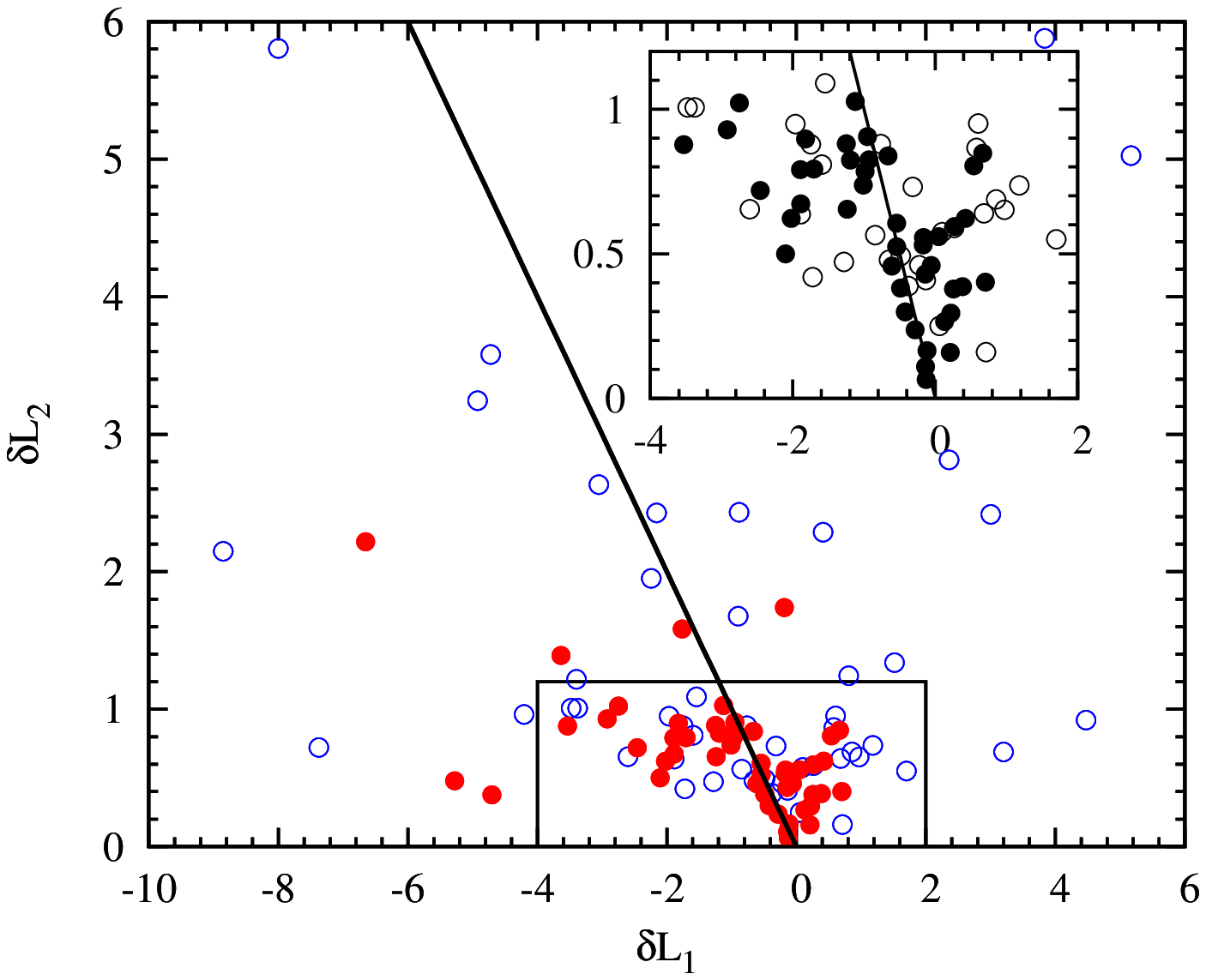}
\caption{Excess in luminosity of the secondary component ($\delta L_2 = L_{2}-M_{2}^{4} $) is 
plotted with respect to luminosity lack (or excess) of the primary components 
($\delta L_1 = L_{1}-M_{1}^{4}$). If  the energy transfer is the 
dominant operating mechanism, all the CBs should make a sequence on the $\delta L_2 = -\delta L_1$ line (solid line).
This is the case for only some of W-subtype CBs. 
}
\end{figure}

The energy transfer process does not work, at least,  for the A-subtype CBs. For the W-subtype CBs, however, 
there are some systems (about 12) 
in which the transfer process may play a part for very close effective temperatures. For some binaries, however, 
both components have excess luminosity which the energy transfer process cannot explain.

%\section{How does initial mass depend on luminosity excess?}
\section{Method for computation of initial mass from luminosity excess}
%\section{Method derived from the MESA Models} 
As one of the essential parameters of a star, luminosity depends on the nuclear burning in the core. 
For a single star, this is determined by the initial mass of the star.
For a component star which has experienced 
mass transfer (and loss),
\begin{equation}
M_{\rm i}=M_{\rm f}+\Delta M,
\end{equation}
where $ \Delta M $ is the mass lost or transferred from the secondary, i.e. the mass difference 
between the initial ($M_{\rm i}$) and the final ($M_{\rm f}$) masses.   
The physical conditions in the central regions of the secondary star,
which yield luminosity higher than $M_{\rm f}^4$, are similar to that of a star with 
a mass lower than $M_{\rm i}$ but higher than $M_{\rm f}$. More precisely, the central 
physical conditions of the secondary star are very close to 
a single star of mass $M_L\approx L^{0.25}$ (in solar units). For $M_L$, which is the mass of an isolated star with 
the same luminosity, 
one can use a more realistic mass-luminosity relation. 
The $M-L$ relation we employ is 
\begin{equation}
L=1.49 M^{4.216},  
\end{equation}
derived from the grids obtained by \citet{yldz} 
for TAMS models. 
The value of mass according to luminosity strongly depends on
the evolutionary phase of a mass losing star at the beginning of  the mass loss/transfer process.
For the computations relating to the 
secondary stars of CBs, for now, we assume that mass transfer starts
near the TAMS phase of the old massive component. Therefore, equation (2) is used throughout this paper
[see Section 5.3 for correction due to deviation from equation (2)].
For a typical TAMS star, the quantity $M_L$ defined by $M_L=(L_{2}/1.49)^{1/4.216}$ for the secondary star computed 
from equation (2) is equal
to the observed $M_2=M_{\rm f}$. If there is a luminosity excess, then  $M_L > M_2$. We define the mass 
difference between these two masses as $\delta M$:
\begin{equation}
\delta M=M_L- M_2. 
\end{equation}
 
According to our assumption, the luminosity excess of the secondary component of a CB is related 
to its initial mass as much as to
its present mass,  and therefore to the  
amount of total mass lost by the secondary ($ \Delta M $). This means that there must be a correlation between 
$\Delta M$ and $\delta M$: $\Delta M= f(\delta M$). 
Then the initial mass of the secondary component of a CB is given as
\begin{equation}
M_{\rm 2i}=M_{2}+f(\delta M)
\end{equation}
We note that $M_{\rm 2i}$
is a function only of the observable quantities $L_2$ and $M_2$. 
The masses, luminosities and radii are in solar units. 

If one finds $M_{\rm 2i}$, then 
%it is straightforward to 
$M_{\rm 1i}$ can be computed by using the 
constraint on the reciprocal of initial mass ratio $1/q_{\rm i}$: $0 < 1/q_{\rm i} < 1$.
$M_{\rm 1i}$ is a function of $\Delta M=M_{\rm 2i}-M_2$ and mass lost by the system ($M_{\rm lost}$).
The increase in the primary mass can be written as
\begin{equation}
M_{\rm 1}-M_{1i} = \Delta M - M_{\rm lost}= \Delta M (1-\gamma).
\end{equation}
where $\gamma$ is the ratio $M_{\rm lost}/\Delta M$. $\gamma=0$ corresponds to the conservative case and 
$\gamma=1$ is the case in which all the mass lost by secondary component is lost by the system. 
The reciprocal of the initial mass ratio as a function of current masses, $M_{\rm 2i}$, and $\gamma$ is given 
as 
\begin{equation}
1/q_{\rm i}=\frac{M_{\rm 1i}}{M_{\rm 2i}}=\frac{M_1-(M_{\rm 2i}-M_2)(1-\gamma)}{M_{\rm 2i}}.
\end{equation}
The unknown value of $\gamma$ can be found from the constraint $0 < 1/q_{\rm i} < 1$ (see section 4.2).
Our next task is to find an expression for $f(\delta M)$.

\subsection{Method derived from the \textsc{mesa models}}
\subsubsection{Initial masses of the secondary components}

As discussed above, the secondary components of CBs have a completely different 
interior from the normal MS stars 
because their initial mass is significantly higher than the present mass. 
This situation forces the secondary components to have a different 
evolutionary path from single MS stars without mass-loss. 
Change in mass essentially causes a breakdown in hydrostatic equilibrium,
which is then restored by expansion of the central regions. 
We construct models in order to assess the ultimate effect of 
change in mass on the structure of stars by using the \textsc{mesa} stellar evolution code. 
From these models, we aim to find an expression for $\Delta M$ as a function of $\delta M$.

Models with constant and variable masses are constructed by using the \textsc{mesa} code. The details of this 
code can be found in \cite{bb13}. The chemical composition and convective parameter for all the models 
are taken as X=0.700, Z=0.02, $\alpha$=1.89. Stellar models are constructed with the solar mixture given by
\cite{grevesse}.
The radiative opacity is derived from
recent OPAL tables \citep{iglesias}, implemented by the low-temperature tables of 
\cite{ferguson}.
The thermonuclear reactions rates are computed by using \cite{angulo}
and \cite{caughlan}, and the standard mixing length theory is employed for convection \citep{cox}.
The \textsc{mesa} $\rho-T$  tables are based on the 2005 update of the
OPAL EOS tables \citep{rogers}.

The progenitors of the secondary stars of CBs should be evolved stars (Hilditch et al. 1988), otherwise 
an excess in luminosity cannot be observed.
Therefore, when we construct our stellar
models we assume that mass transfer begins near the TAMS which we take
to be when the central hydrogen abundance is $10^{-6}$.
The basic properties of the constructed models with different initial masses are listed in Table 1. 
The range of initial masses is $1.2-3.5\mathrm{M}_\odot$.
For all of these models, the mass loss rate is 
$10^{-8}\mathrm{M}_\odot\mathrm{yr^{-1}}$ and the final mass is $0.5\mathrm{M}_\odot$.
After several tries for different mass-loss rates, we deduce that the results are independent of the mass-loss rate.
The corresponding mass to the luminosity ($M_L$) after mass transfer is computed from equation (2). 
The mass difference $\delta M$ is as given in equation (3) and  $\Delta M=M_{\rm i}-M_{\rm f}$.
We also construct models with $M_{\rm f}=0.8\mathrm{M}_\odot$. The basic properties of these models are given in Table 2. 
\begin{table}
\caption{Model results with mass-loss to obtain an expression for the difference between initial and final masses. 
The final mass is chosen as $ 0.5\mathrm{M}_{\odot} $ and  the mass transfer begins when central hydrogen abundance $ X_{\rm C}=10^{-6} $. }
\centering
\small\addtolength{\tabcolsep}{-3pt}
\begin{tabular}{ccccccc}
\hline
Model	&	$M_{\rm i}$	&	$M_{\rm f}$	&	$L_{\rm f}$	&	$M_{L}$	&	$\delta M$	&	$\Delta M$	\\
No	&	$(\mathrm{M}_{\odot})$	&	$(\mathrm{M}_{\odot})$	&	$(\mathrm{L}_{\odot})$	&	$(\mathrm{M}_{\odot})$	&	$(\mathrm{M}_{\odot})$	&	$(\mathrm{M}_{\odot})$	\\
\hline\hline
1	&	1.20	&	0.50	&	0.251	&	0.708	&	0.208	&	0.700	\\
2	&	1.50	&	0.50	&	0.359	&	0.774	&	0.274	&	1.000	\\
3	&	1.80	&	0.50	&	0.638	&	0.894	&	0.394	&	1.300	\\
4	&	2.50	&	0.50	&	2.884	&	1.303	&	0.803	&	2.000	\\
5	&	3.00	&	0.50	&	6.457	&	1.594	&	1.094	&	2.500	\\
6	&	3.50	&	0.50	&	11.220	&	1.830	&	1.330	&	3.000	\\
\hline
\end{tabular}
\end{table}
\begin{table}
\caption{ Same as Table 1, but for the final mass  $M_{\rm f}=0.8\mathrm{M}_{\odot}$. 
}
\centering
\small\addtolength{\tabcolsep}{-3pt}
\begin{tabular}{ccccccc}
\hline
Model	&	$M_{\rm i}$	&	$M_{\rm f}$	&	$L_{\rm f}$	&	$M_{L}$	&	$\delta M$	&	$\Delta M$	\\
No	&	$(\mathrm{M}_{\odot})$	&	$(\mathrm{M}_{\odot})$	&	$(\mathrm{L}_{\odot})$	&	$(\mathrm{M}_{\odot})$	&	$(\mathrm{M}_{\odot})$	&	$(\mathrm{M}_{\odot})$	\\
\hline\hline
1	&	1.20	&	0.80	&	0.893	&	0.972	&	0.172	&	0.400	\\
2	&	1.50	&	0.80	&	1.064	&	1.016	&	0.216	&	0.700	\\
3	&	1.80	&	0.80	&	1.346	&	1.077	&	0.277	&	1.000	\\
4	&	2.50	&	0.80	&	3.926	&	1.408	&	0.608	&	1.700	\\
5	&	3.00	&	0.80	&	8.630	&	1.714	&	0.914	&	2.200	\\
6	&	3.50	&	0.80	&	17.378	&	2.042	&	1.242	&	2.700	\\
\hline
\end{tabular}
\end{table} 

The initial mass is plotted with respect to $M_L$ in Fig. 3 for $M_{\rm f}=0.5\mathrm{M}_\odot$ and $0.8\mathrm{M}_\odot$.
The relation between the initial and luminosity masses is
almost linear.
In both sets of models with either $M_{\rm f}=0.5\mathrm{M}_\odot$ or $0.8\mathrm{M}_\odot$,
we note that the higher the $\Delta M$, the higher is the $\delta M$. In Fig. 4, 
$\Delta M$ is plotted with respect to
$\delta M$ (in units of $\mathrm{M}_\odot$). Both sets of models with $M_{\rm f}=0.5\mathrm{M}_\odot$ (filled circle) 
and $M_{\rm f}=0.8\mathrm{M}_\odot$ (circle) have the same relation between 
$\Delta M$ and $\delta M$. The solid line shows the fitted line 
$[2.50 (\delta M-0.07)^{0.64}]$
for both data sets. It is remarkable to confirm that
the relation is not a linear one. Also shown in Fig. 4 is the dotted line for the linear relationship 
$\Delta M= 2.0 \delta M$.
\begin{figure}
\includegraphics[width=67mm, angle=270]{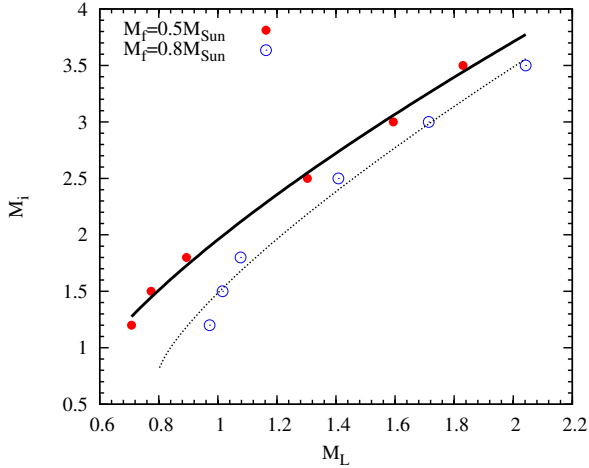}
\caption{The relation between $ M_{L} $ and $ M_{\rm i} $ for the final masses $ 0.5$ (filled circle) and  $ 0.8$ (circle) $\mathrm{M}_{\odot} $. 
}
\end{figure}
%\begin{figure}
%\includegraphics[width=67mm, angle=270]{Fig_9_ML_vs_M2i_Mf0.8_u4.ps}
%\caption{The relation between $ M_{L} $ and $ M_{\rm i} $ for the final mass $ 0.8$ M$_{\odot} $. 
%}
%\end{figure}
\begin{figure}
\includegraphics[width=67mm, angle=270]{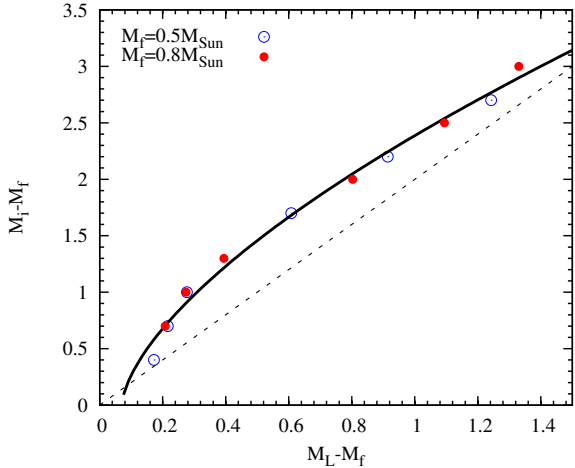}
\caption{$\Delta M$ as a function of $\delta M$ for the \textsc{mesa} models with final masses 
$M_{\rm f}=0.5$ (filled circle) and $M_{\rm f}=0.8\mathrm{M}_\odot$ (circle). 
The solid line represents the fitting curve
$\Delta M= 2.50 (\delta M-0.07)^{0.64}$ and dotted line is for $\Delta M= 2.0 \delta M$.
}
\end{figure}

The initial mass of the secondary stars of CBs is then computed by using   
\begin{equation}
M_{\rm 2i}=M_2+\Delta M=M_2+2.50 (\delta M-0.07)^{0.64}
\end{equation}
where the masses are in units of solar mass. 
 
\subsubsection{Uncertainties in the initial masses of the secondary components}
The initial masses of the secondary components are computed from $L_{2}$ and $M_{2}$.
Therefore, the uncertainty in $M_{\rm 2i}$ ($\Delta M_{\rm 2i}$) arises from uncertainties $\Delta M_2$ and $\Delta L_2$.
Define $f=\Delta M=2.50(\delta M- 0.07)^{0.64}$. Then, $f$ is a function of $M_L$ and $M_2$.
In terms of these quantities and their uncertainties, the uncertainty in $M_{\rm 2i}$ can be written as
\begin{equation}
\Delta M_{\rm 2i} = \Delta M_2+\frac{\partial f}{\partial M_L}\frac{d M_L}{d L_2}\Delta L_2
                          +\left|\frac{\partial f}{\partial M_2}\right| \Delta M_2.
\end{equation}
Using the definition of $\delta M$ given in equation (3),
we find that
\begin{equation}
\frac{\partial f}{\partial M_L}=\left| \frac{\partial f}{\partial M_2}\right|=1.59 (\delta M- 0.07)^{-0.36}.
\end{equation}
From the employed $M-L$ relation, we obtain
\begin{equation}
\frac{d M_L}{d L_2}=\frac{0.22}{L_2^{0.76}}.
\end{equation}
The explicit form of $\Delta M_{\rm 2i}$ as a function of $\Delta M_2$ and $\Delta L_2$ is
\begin{equation}
\Delta M_{\rm 2i} = \Delta M_2+\frac{1.59}{(\delta M-0.07)^{0.36}}\left(\frac{0.22 ~ \Delta L_2 }{L_2^{0.76}}+\Delta M_2\right).
\end{equation}

\section{Results and discussion}
\subsection{The initial masses of the secondary components}

The initial mass of the secondary component of a CB is computed using the method from the previous section.
The results are listed in %the twelfth column of 
Table A1.
If we compare an A-subtype  with a 
W-subtype CB of the same $M_2$ (present secondary mass), 
then we find that the initial mass of the secondary of an A-subtype is $0.5 \mathrm{M}_\odot$ higher than  that of W-subtype.
This means that the (present) secondary components of the A-subtype have lost/transferred much more mass
than those of the W-subtype. 
\begin{figure}
\begin{center}
\includegraphics[width=67mm, angle=270]{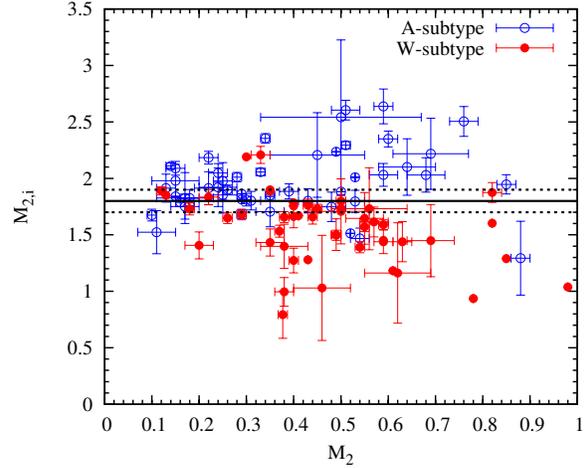}
\caption{Initial mass of the secondary components of CBs. Circles  are for A-subtype and the filled circles are for W-subtype. The solid black line represents the derived transition
mass with the dashed lines showing the $1\sigma$ error bars.
Apart from a few systems, the A-subtype CBs have $M_{\rm 2i}$ higher than $1.80$ $\mathrm{M}_{\odot}$ and  the W-subtype 
CBs have $M_{\rm 2i}$ lower than $1.80\mathrm{M}_{\odot}$.
}
\end{center}
\end{figure}

In Fig. 5, the initial mass of the secondaries, derived by using
the method based on \textsc{mesa} models, is plotted with respect to the present mass of the secondary components. 
The initial masses are plotted with their uncertainties computed by a method presented in the previous section. 
It can clearly be seen that subtypes are determined by the initial mass of the secondaries. 
If $ M_{\rm 2i}>1.8\pm0.1\mathrm{M}_{\odot} $, the system becomes A-subtype. If  $ M_{\rm 2i}<1.8\pm0.1\mathrm{M}_{\odot} $, 
then the system exhibits the W-subtype properties. 
There are just a few systems that do not obey this prescription (see below). 
The mean initial mass of the secondaries 
is $\overline{M}_{\rm 2iA} = 1.97\mathrm{M}_\odot$  
for the A- and $\overline{M}_{\rm 2iW} =1.56\mathrm{M}_\odot$ for the W-subtype CBs. 
The underlying mechanism for this separation must be based 
on competition between the orbital and the nuclear time-scales.

There are only two W-subtype systems (Wa) having an initial mass higher than $\overline{M}_{\rm 2iA}$.
They are V728 Her and V402 Aur.
These binaries are the hottest of the W-subtype CBs. Their $M-L$ relation and angular momentum properties
are similar to that of the A-subtype rather than the W-subtype. 
The number of the A-subtype binaries (Aw) having an initial mass significantly less than $\overline{M}_{\rm 2iW}$, 
however, is four.
These binaries are EQ Tau, OO Aql, V508 Oph and TZ Boo.
They have $T_{\rm e}$ about 5600 K $[\log(T_{\rm e})=3.75]$ and are the coolest of the A-subtype CBs. 
The O'Connell effect may be responsible for
this atypical situation if it is not due to uncertainties
in the observed
$T_{\rm e}$s.
However, in addition to 
the initial mass of the secondary components, another
secondary effect may also play a role in the formation of W UMa subtypes. For example,
initial orbital angular momentum may be important.

Putting the atypical CBs aside, the initial mass range for the secondary components 
is $1.7-2.6 \mathrm{M}_{\odot}$ for A-subtype and $0.7-1.9 \mathrm{M}_{\odot}$ for W-subtype. However, for most of the CBs
the range is $1.3-2.2 \mathrm{M}_{\odot}$.

CC Com is the system with the minimum secondary initial mass, 0.86 $\mathrm{M}_{\odot}$. 
The nuclear time-scale 
of such a star is about 16 Gyr, greater than the age of the galaxy.
Therefore, this system (or most of the W-subtype CBs) must have some specific feature, 
which makes its fundamental properties unusual.
As a matter of fact, both components of CC Com are brighter than ZAMS stars and fainter than TAMS stars. 
Formation of binaries with $M_{\rm 2i} < 1.25\mathrm{M}_{\odot}$ seems to be different from the other CBs. 
They have either a different value
of $\gamma$ or different structure for the progenitors of the secondary components during the
mass transfer process. For example, the progenitors may not be as evolved (not TAMS) as  we assume 
(see Section 5.3).

The most realistic uncertainties in $M_{\rm 2i}$ are computed from equation (11) and given in Table A1.
For some of the binaries, for example, HV UMa, the uncertainty $\Delta M_{\rm 2i}$ is 
very high due to high uncertainties in $M_2$ and $L_2$.
However, this result does not change the situation much because the uncertainties are
within an acceptable range for most of the binaries in our sample. 

\subsection{Initial masses of the primary components}
\begin{figure}
\includegraphics[width=67mm,angle=270]{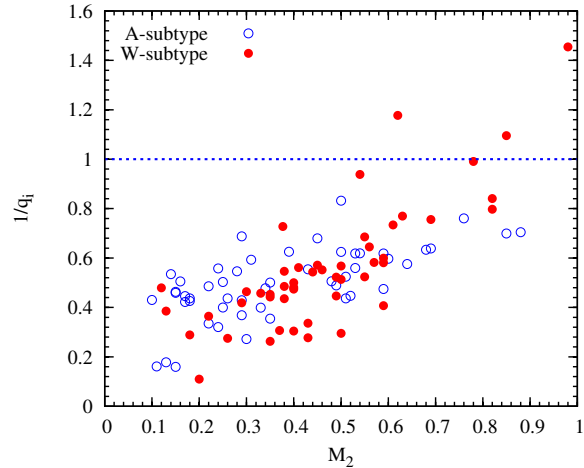}
\caption{$ 1/q_{\rm i} $ computed from the expressions derived from
the models by using \textsc{mesa} code is plotted with respect to $M_2$. 
All the A-subtype systems are in the range $ 0<1/q_{\rm i}<1 $; 
just three W-subtype systems are greater than one.
}
\end{figure}
The reciprocal mass ratio $1/q_{\rm i}$ is the physical constraint for the determination of initial 
masses of the primary components. %, as in Section 3.2. 
The fitting parameter $\gamma$ (equation 5) is the mass lost by the
system in units of $\Delta M$. 
First, we consider A-subtype CBs. If $\gamma$ has a small value, then  some A-subtype systems have an unreasonably negative value for $M_{\rm 1i}$.
In order to keep  $M_{\rm 1i}>0$ ($1/q_{\rm i} >0$) for all A-subtype CBs, the minimum value of $\gamma$ is $0.50$.
The maximum value of $\gamma$ for $1/q_{\rm i} < 1$, however, is $0.87$.  
These two values of $\gamma$ give mean value
for $1/q_i$ as $0.368$ and $0.666$, respectively. For $1/\overline{q}_i=0.5$,  $\gamma=0.664$.
This value of  $\gamma$ yields $M_{\rm 1i}$ ranging from $0.3$ to $1.5\mathrm{M}_\odot$ for A-subtypes. This means that the mass lost by these subtypes is higher than the transferred mass: 66 per cent of mass lost by the secondary components is lost by the system and 34 per cent is transferred to the primary components.

$1/q_{\rm i}$ of A-subtype CBs with $\gamma=0.664$ is plotted with respect to $M_2$ in Fig. 6.
The reciprocal of the initial mass ratio is well within the expected range. 
For $1/q_{\rm i}$ of W-subtype CBs, we use the same value of $\gamma$.
On average, the A-subtype CBs have higher $1/q_{\rm i}$ than the W-subtype CBs.  
There are only 3 violating systems.
These binaries are the W-subtype CBs (see below) and have very low $\delta M$ values, less than $0.2\mathrm{M}_{\odot}$.  %unutma
They  may not be TAMS stars but are somehow in between the ZAMS and the TAMS lines.

The value of $\gamma$ we found is a  mean value. Indeed, there is no restriction for a single value of $\gamma$
for all CBs. However,
what is important is that there is a solution with a single value of $\gamma$ for almost all 
CBs, and this is certainly true for the A-subtype CBs. It may be concluded that
the value of $\gamma$ does not dramatically change much from system to system. 

The initial masses of the primary components are listed in %the thirteenth column of 
Table A1.
The range of primary initial mass of the CBs is $0.2-1.5\mathrm{M}_{\odot}$, apart from two binaries 
(HV UMa and V376 And). This is the mass range in which magnetized stellar wind is an efficient
mechanism for angular momentum loss \citep{tut32}. 
The mean value of the initial primary masses is $\overline{M}_{\rm 1i} = 0.99 \mathrm{M}_\odot$ 
for the A- and $\overline{M}_{\rm 1i} = 0.84 \mathrm{M}_\odot$ for the W-subtype.
The uncertainties in $M_{\rm 1i}$ are also given in Table A1.
$\Delta M_{\rm 1i}$ is taken as $\Delta M_{\rm 2i}/q_{\rm i}$.

\subsection{Correction of the initial masses of components of CBs with $\delta M < 0.35\mathrm{M}_\odot$}
\begin{figure}
\includegraphics[width=67mm,angle=270]{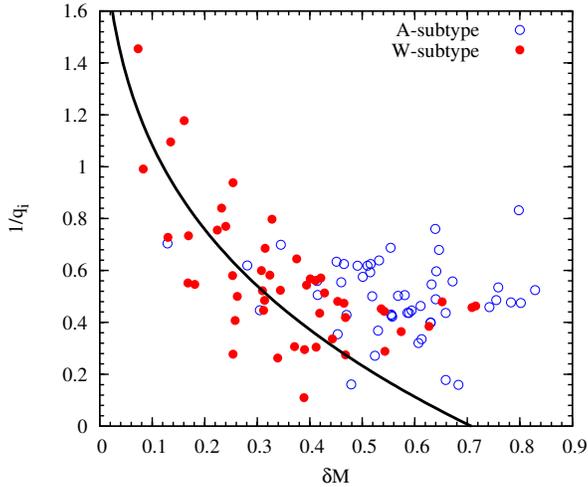}
\caption{$ 1/q_{\rm i} $ computed from the expressions derived from
the models by using the \textsc{mesa} code is plotted with respect to $\delta M$. 
$ 1/q_{\rm i} $ for the A-subtype CBs has a very uniform distribution. For
the W-subtype CBs, however, this is not the case. For small values of $\delta M$, $ 1/q_{\rm i} $ is dependent on 
$\delta M$.
The solid line shows the fitted curve for the W-subtype CBs ($-3.05 \delta M^{0.25}+2.80$).
}
\end{figure}
\begin{table}
\caption{The corrected masses for the secondary and the primary masses of the W-subtype CBs 
with $\delta M<0.35\mathrm{M}_\odot$. The correction is computed according to the expression 
obtained from the curve in Fig. 7 ($-3.05 \delta M^{0.25}+2.80$).
}
\centering
\small\addtolength{\tabcolsep}{-3pt}
\begin{tabular}{lcccccc}
\hline
Star & $M_{1}$ &       $M_{2}$ & $M_{\rm 2i}$ &       $M_{\rm 1i}$ &       $1/q_{\rm i}$     & $t_{\rm MS}(Gyr)$\\
\hline
DN      Cam &   1.85 &   0.82 &   1.86 &   1.50 &   0.81 &   1.25 \\
EF      Boo &   1.61 &   0.82 &   1.90 &   1.25 &   0.66 &   1.17 \\
ET    Leo   &   1.59 &   0.54 &   1.55 &   1.25 &   0.81 &   2.19 \\
AA      UMa &   1.56 &   0.85 &   1.88 &   1.21 &   0.65 &   1.20 \\
YY      Eri &   1.55 &   0.62 &   1.53 &   1.24 &   0.81 &   2.28 \\
ER      Ori &   1.53 &   0.98 &   1.74 &   1.27 &   0.73 &   1.52 \\
BB    Peg   &   1.42 &   0.55 &   1.59 &   1.07 &   0.67 &   2.03 \\
AE      Phe &   1.38 &   0.63 &   1.69 &   1.02 &   0.61 &   1.68 \\
W       UMa &   1.35 &   0.69 &   1.77 &   0.99 &   0.56 &   1.44 \\
AM  Leo     &   1.29 &   0.59 &   1.64 &   0.94 &   0.57 &   1.85 \\
V781    Tau &   1.29 &   0.57 &   1.61 &   0.94 &   0.59 &   1.96 \\
RZ      Com &   1.23 &   0.55 &   1.58 &   0.89 &   0.56 &   2.09 \\
AO      Cam &   1.12 &   0.49 &   1.54 &   0.77 &   0.50 &   2.26 \\
GZ      And &   1.12 &   0.59 &   1.68 &   0.75 &   0.45 &   1.69 \\
TW      Cet &   1.06 &   0.61 &   1.74 &   0.68 &   0.39 &   1.52 \\
BX      Peg &   1.02 &   0.38 &   1.42 &   0.67 &   0.47 &   2.90 \\
AB      And &   1.01 &   0.49 &   1.54 &   0.66 &   0.43 &   2.25 \\
SW      Lac &   0.98 &   0.78 &   1.88 &   0.61 &   0.33 &   1.21 \\
VW      Cep &   0.93 &   0.40 &   1.48 &   0.57 &   0.38 &   2.56 \\
V757    Cen &   0.88 &   0.59 &   1.74 &   0.49 &   0.28 &   1.52 \\
CC      Com &   0.72 &   0.38 &   1.49 &   0.35 &   0.23 &   2.50 \\
XY      Leo &   0.76 &   0.46 &   1.68 &   0.35 &   0.21 &   1.69 \\
V523    Cas &   0.75 &   0.38 &   1.53 &   0.36 &   0.24 &   2.28 \\
BH      Cas &   0.74 &   0.35 &   1.36 &   0.40 &   0.29 &   3.32 \\
RW      Dor &   0.64 &   0.43 &   1.63 &   0.24 &   0.14 &   1.86 \\
\hline
\end{tabular}
\end{table} 
As previously stated, the initial mass of some secondaries is very small for some CBs,
for example CC Com.
In Fig. 7, the reciprocal of initial mass ratio $1/q_{\rm i}$ is plotted with respect to $\delta M $ both
for the A- and the W-subtype CBs. For A-subtype, $1/q_{\rm i}$ does not show any dependence on  $\delta M $.
For W-subtype, however, $1/q_{\rm i}$ is dependent on $\delta M $ if $\delta M < 0.4\mathrm{M}_\odot$. 
The solid line shows the curve fit 
$(-3.05 \delta M^{0.25}+2.80)$ for this range. 
This relation reflects our assumption about the evolutionary phase of secondary 
components at the beginning of  mass transfer. It seems that for some of the W-subtype binaries 
the secondary components are not close to the TAMS line. Therefore, equation (7) is not valid for these 
CBs. Using the curve $(-3.05 \delta M^{0.25}+2.80)$ of Fig. 7, we obtain a new equation for $ M_{\rm 2i}$, and 
$ M_{\rm 1i}$ is computed as in Section 5.2.
This equation removes the slope in Fig. 7. 

The corrected initial masses of the secondary and the primary 
components of these 
CBs are listed in Table 3. 
The upper limit for the initial mass of the secondaries is $1.90\mathrm{M}_{\odot}$, consistent with the results given above. 
The minimum initial mass of the secondaries is 
with small $\delta M$ pertaining to BH Cas, $1.36\mathrm{M}_{\odot}$. 

\subsection{Some notes on evolutionary time-scales and cluster member CBs}
As mentioned above, the ultimate fate of a binary depends on its total mass and 
how this mass is shared between 
its components. If mass of the massive component ($M_{\rm 2i}$) is high enough, then nuclear evolution is very fast compared to orbital evolution. 
Therefore, the lifetime of this star is so short that the system cannot be a CB. 
In this respect, a very interesting result we obtain is that there is an upper limit for the mass
of the massive components of CBs, $2.61 \mathrm{M}_{\odot}$ (V921 Her). 
This result leads us to consider binaries with components
having a mass higher than $2.61 \mathrm{M}_{\odot}$ as candidate systems for binaries with compact objects, for example.

Ages of the W UMa type CBs are investigated by many different groups using different methods 
\citep{bg1,VV1,bb05,LH1,LEA1} and range from $0.1-1.0$ to $7.2 \mathrm{Gyr}$.
The life-time of a W UMa type CB can roughly be divided into three different phases.
These are detached, semi-detached and contact phases. From the initial masses, at least for the A-subtype CBs,
we find how long the detached phase is: $t_{\rm det}=t_{\rm MS}$.
The MS age for a star with $2.61\mathrm{M}_{\odot}$ is $0.46 \mathrm{Gyr}$. This time is likely to be very short for angular momentum evolution 
without mass-transfer. The MS age of a star with transition mass $(1.8\mathrm{M}_{\odot})$ from the A- to the W-subtype CBs
is $1.37 \mathrm{Gyr}$ (EX Leo). For such systems, the nuclear and angular momentum time-scales are comparable.
For some W-subtype systems, the nuclear time scale is liable to be so long that the components evolve into contact phase before
significant nuclear evolution of the 
massive component. The MS lifetime for BH Cas, for example is $3.32 \mathrm{Gyr}$.
Such a time is very long for angular momentum evolution of BH Cas and similar CBs
with two late-type components. 

\begin{table}
\caption{Cluster member CBs. 
}
\centering
\small\addtolength{\tabcolsep}{-3pt}
\begin{tabular}{lcccccccl}
\hline
Star   & $M_{\rm 1i}$ & $M_{\rm 2i}$ & $t_{\rm MS}(\mathrm{Gyr})$ & $t_{\rm cl}(\mathrm{Gyr})$ & P(day)& {Subtype} & Cluster \\
\hline
TX Cnc  & 0.50 & 1.70  & 1.6  &  0.7  &  0.3830 &  W  & Praesepe \\ % 0.50      -0.9
AH Vir  & 0.94 & 1.67  & 1.7  &  3.0  &  0.4075 &  W  & Wolf 630  \\% 0.94       1.3
QX And  & 0.62 & 1.92  & 1.1  &  2.0  &  0.4118 &  A  & NGC 752 \\   % 0.62      0.9
AH Cnc  & 0.93 & 1.85 &  1.3  &  4.0  &  0.3604 &  A  & M67\\       %   0.93     2.7
\hline
\end{tabular}
\end{table} 
According to the results,  we deduce that ages of the A-subtype binaries $(t_s)$ are greater than 
the MS lifetime of
the progenitor of their secondaries $(t_{\rm MS})$. For the W-subtype binaries, however, $t_s$ may also be less than
$t_{\rm MS}$ because it seems that the mass transfer for some CBs starts before the secondary component $(M_{\rm 2i})$
reaches the TAMS line. In order to test this point, we consider CBs in clusters.
There are four such binaries in our CB sample. Two of them are the W-subtype CBs. 
 The W-subtype binaries TX Cnc and AH Vir are
members of Praesepe and Wolf 630, respectively. The A-subtype CBs are QX And of  NGC 752 and
AH Cnc of M67. %The latter system is a system with third body. 
The MS lifetime according to $M_{\rm 2i}$ $(t_{\rm MS})$
of the binaries and age of clusters are listed in Table 4, respectively.
While $t_{\rm MS}$ is less than $t_{\rm s}=t_{\rm clus}$ (cluster age) for A-subtype,
$t_{\rm MS}$ for the W-subtype binaries is less than $t_{\rm s}$ for AH Vir and greater than for TX Cnc. 
Although there are only four well-known cluster member systems,  these results are consistent with
our expectations.

\begin{table}
\caption{Mass intervals for final products. }
\centering
\small\addtolength{\tabcolsep}{-3pt}
%\begin{tabular}{ccccl}
\begin{tabular}{ccccl}
\hline
$\mathrm{min}M_{\rm 2i}(\mathrm{M}_\odot$) & $\mathrm{max}M_{\rm 2i}$ $(\mathrm{M}_\odot)$ & $M_{\rm 1i}$ $(\mathrm{M}_\odot)$   & $t_{\rm min}(\mathrm{Gyr})$& Product     \\
\hline
1.3$\pm$0.1  & 1.8$\pm$0.1  &0.2-1.5 & 1.4  &  W-subtype \\
1.8$\pm$0.1  & 2.6$\pm$0.1  &0.2-1.5 & 0.5-1.4  &  A-subtype \\

\hline
\end{tabular}
\end{table} 
Table 5 summarizes the results obtained for the initial masses of the A- and the W-subtype CBs. 
The MS lifetime according to $M_{\rm 2i}$ is given in the fourth column. %This is the minimum age for the-A subtype.
The age of the A-subtypes must be greater than this age.
For the W-subtype CBs, however, age can be either greater or smaller than the given value $1.4 \mathrm{Gyr}$,
depending on the efficiency of the angular momentum loss mechanism via magnetic stellar wind.  
 
\section{Conclusion}
In the present study we have investigated the physical properties of CBs. 
The approximate initial masses of  the secondary and primary 
components are computed, for the first time ever, by using 
the basic parameters of 100 W UMa type CBs (51  A- and 49 W-subtype).
The initial mass of the secondary is computed from its luminosity excess by a method
based on the expression derived from stellar models with mass-loss.
We have also deduced that only one-third of the mass lost by the secondary components
is transferred to the primary components. The remaining part leaves the systems.
The range of initial masses of the secondaries of CBs is $1.3-2.6 \mathrm{M}_{\odot}$.
The upper part of this range is occupied by the A-subtype CBs and the W-subtype CBs are in the 
lower part.
The transition occurs at about $1.8\pm 0.1\mathrm{M}_\odot$. 
%This result leads us to confirm that
%detached binary systems with an early-type secondary (the component that was {\it more} massive before  
%%the mass transfer process) of mass $M_{\rm 2i} <$ 2.6 M$_\odot$ become  A-subtype CBs, while the W-subtype
%CBs are mostly formed from the detached binaries with late-type components.
%\par\noindent ****** I didn't understand your last sentence of this paragraph,  but it seems to me to be
%only saying in a different form what you have already said. *********

The main assumption in our method is that  the mass transfer starts near or after the TAMS phase
of the massive component (progenitor of the secondary component). This assumption seems fairly good for 
the A-subtypes. For the W-subtype, however, a correction is needed. From $\delta M$ and the reciprocal mass-ratio diagram 
(Fig. 7), we develop a method and obtain corrected initial masses for the case of
small $\delta M$. The corrected initial masses of the secondaries range from $1.3-1.9 \mathrm{M}_{\odot}$.
One of the important outcomes of the present study is that there is also a lower limit for the 
initial mass of the primary components. The reason for this result
should be studied in the context of angular momentum evolution versus nuclear evolution. 
 
For the binary systems with an initial mass higher than $1.8 \mathrm{M}_{\odot}$, although the angular momentum is 
lost by the less massive component the nuclear evolution is principally the mechanism 
responsible for the Roche lobe filling process. Therefore, semi-detached systems with $M_{\rm 2i}=1.8-2.61 \mathrm{M}_{\odot}$ must yield 
A-subtype W UMa CBs. The binary systems with $M_{\rm 2i}$ less than $1.8 \mathrm{M}_{\odot}$ evolve into the contact phase due to 
the rapid angular momentum evolution. The Roche lobes of their components contract so that they form
the W-subtype contact or near-contact binaries despite the components not being much evolved in the MS phase.
 
The primary components of W UMa type CBs have an initial mass range from $0.2$ to $1.5 \mathrm{M}_{\odot}$. 
%\par\noindent ******* Do you mean the initial version of the current primary, i.e the component that was 
%initially the secondary system? I find this terminology rather confusing, and I suspect others will.
%Although it's a bit late, I would favour A and B as the names of the {\it currently} more and less
%massive components, and $*1$ and $*2$, pronounced `star 1' and `star 2', as the names of the {\it initially} 
%more and less massive components. ************
%
%\par\noindent 
A star with mass in this range loses its spin angular momentum relatively rapidly. 
If it has a tidally interacting companion,
orbital angular momentum is lost by the system. The precursors of W UMa type CBs are the systems 
in which nuclear evolution of the massive components is in competition with the angular momentum loss from
the less massive components. In the precursors of the W-subtype W UMa CBs, both components are effective 
in the angular momentum loss process.

The method we develop for computation of the initial masses of the secondary components is highly suitable
for a complete error analysis. From the derivatives of the function derived from the fitting curve for 
$\Delta M$ as a function of $L_2$ and $M_2$ (equation 8), we successfully compute $\Delta M_{\rm 2i}$ in terms
of the observed uncertainties $\Delta L_2$ and $\Delta M_2$ (equation 11).

We also consider cluster member W UMa type CBs. Although the number of CBs with accurate dimensions available 
in the literature is very small (4 systems), the results support our findings on initial masses and time-scales for the A- and W-subtype CBs. 

\section*{Acknowledgments}

The anonymous referee and Peter P. Eggleton are  acknowledged for 
their suggestions which improved presentation of the manuscript.

\newpage
\onecolumn
\appendix
\section[]{Basic properties of W UMa  type binaries}
%\afterpage{\input{Table1}}
%\afterpage{\clearpage\input{Table1}}
%\twocolumn
%\begin{center}
%\begin{sidewaystable}
%\begin{landscape}
\small\addtolength{\tabcolsep}{-4pt}
\begin{longtable}{@{}lrcccccccccclll@{}}
%\small
\caption[Continued.]{Fundamental properties of the A- and W-subtype CBs.
Columns are organized as name, subtype, period, mass ratio,
effective temperatures of primary and secondary, primary and secondary radii, luminosities of
primary and secondary, primary and secondary masses, initial masses of primary and secondary before the mass
transfer (see Section 5), reciprocal of initial mass ratio, MS life time of secondary component according
to initial mass and
numbers for references.
%The second row of each CB is devoted to uncertainties if available in the literature.
}\\
\hline
%\small
Star	&t $~$~ $P$ ~$~$~&	$T_{1}$	&	$T_{2}$	&	$R_{1}$	&	$R_{2}$	&	$L_{1}$	&	$L_{2}$	&	$M_{1}$	&	$M_{2}$	&	$M_{\rm 2,i}$	&	$M_{\rm 1,i}$	&	$1/q_{\rm i}$	&	$t_{\tiny \rm MS}$ & Ref	\\
        & d ~$~$~ &		$(K)$	&	$(K)$	&	(R$_{\odot})$	&	(R$_{\odot})$	&	(L$_{\odot})$	&	(L$_{\odot})$	&	(M$_{\odot})$	&	(M$_{\odot})$	&	(M$_{\odot})$	&	(M$_{\odot})$	&	&Gy&	\\    
\hline
\endfirsthead

\caption[-- continued from previous page]{-- continued from previous page}\\
\hline
Star & t $~$~  $P$ ~$~$~&	$T_{1}$	&	$T_{2}$	&	$R_{1}$	&	$R_{2}$	&	$L_{1}$	&	$L_{2}$	&	$M_{1}$	&	$M_{2}$	&	$M_{\rm 2,i}$	&	$M_{\rm 1,i}$	&	$1/q_{\rm i}$	&	$t_{\tiny \rm MS}$ & Ref	\\
           &d ~$~$~&		$(K)$	&	$(K)$	&	(R$_{\odot})$	&	(R$_{\odot})$	&	(L$_{\odot})$	&	(L$_{\odot})$	&	(M$_{\odot})$	&	(M$_{\odot})$	&	(M$_{\odot})$	&	(M$_{\odot})$	&	&	&	\\    
\hline
\endhead

\hline
\endfoot

HV      UMa& A    0.711&     7000$\pm$   200&  7000$\pm$   200&  2.62$\pm$  0.25&  1.18$\pm$  0.16& 17.6 &  4.48$\pm$  1.73&  2.80$\pm$  0.60&  0.50$\pm$  0.17&  2.54$\pm$  0.69&  2.11$\pm$  0.57&  0.83&  0.5& 6,37\\
V376 And& A   0.799&     8350$\pm$   120&  7335$\pm$   120&  2.66$\pm$  0.02&  1.55$\pm$  0.01& 30.4$\pm$  0.43&  6.14$\pm$  0.41&  2.49$\pm$  0.06&  0.76$\pm$  0.03&  2.51$\pm$  0.13&  1.90$\pm$  0.10&  0.76&  0.5&45 \\
RR      Cen& A   0.606&     6918$\pm$   319&  6761$\pm$   312&  2.24$\pm$  0.10&  1.07$\pm$  0.05& 10.2$\pm$  1.65&  2.19$\pm$  0.35&  2.09$\pm$  0.43&  0.45$\pm$  0.10&  2.21$\pm$  0.38&  1.50$\pm$  0.25&  0.68&  0.7&14 \\
V921 Her& A   0.877&     7700$\pm$    20&  7346$\pm$    20&  2.75$\pm$  0.02&  1.41$\pm$  0.01& 23.5$\pm$  0.09&  5.09$\pm$  0.04&  2.07$\pm$  0.05&  0.51$\pm$  0.03&  2.61$\pm$  0.09&  1.37$\pm$  0.05&  0.52&  0.5& 8\\
UZ  Leo& A   0.618&     6980$\pm$    15&  6830$\pm$    15&  2.29$\pm$  0.02&  1.39$\pm$  0.01& 11.0$\pm$  0.18&  3.71$\pm$  0.07&  1.99$\pm$  0.04&  0.60$\pm$  0.02&  2.35$\pm$  0.07&  1.40$\pm$  0.04&  0.60&  0.6&45 \\
V535 Ara& A   0.629&     8200$\pm$   500&  8129$\pm$   521&  2.09$\pm$  0.03&  1.23$\pm$  0.02& 18.0$\pm$  3.00&  6.00$\pm$  1.00&  1.94$\pm$  0.04&  0.59$\pm$  0.02&  2.64$\pm$  0.15&  1.25$\pm$  0.07&  0.47&  0.4& 26\\
AQ      Tuc& A   0.595&     6980 &  6860 &  2.05$\pm$  0.11&  1.32$\pm$  0.07&  8.95$\pm$  0.99&  3.47$\pm$  0.38&  1.93$\pm$  0.21&  0.69$\pm$  0.08&  2.22$\pm$  0.32&  1.42$\pm$  0.20&  0.64&  0.7& 5,13,37\\
EF      Dra& A   0.424&     6000 &  6054 &  1.70 &  0.78 &  3.36 &  0.73 &  1.81 &  0.29 &  1.86 &  1.28 &  0.69&  1.2& 37\\
V2388   Oph& A   0.802&     6900$\pm$    23&  6349$\pm$    23&  2.60$\pm$  0.02&  1.30$\pm$  0.01& 13.5$\pm$  0.20&  2.43$\pm$  0.07&  1.80$\pm$  0.02&  0.34$\pm$  0.01&  2.35$\pm$  0.04&  1.12$\pm$  0.02&  0.48&  0.6& 36\\
AW      UMa& A   0.439&     7175 &  7022 &  1.87$\pm$  0.05&  0.66$\pm$  0.02&  8.30 &  0.95 &  1.79$\pm$  0.14&  0.14$\pm$  0.01&  2.11$\pm$  0.03&  1.13$\pm$  0.02&  0.53&  0.8&27,37 \\
V776 Cas& A   0.440&     6700$\pm$    90&  6725$\pm$    90&  1.82$\pm$  0.02&  0.75$\pm$  0.01&  5.90$\pm$  0.11&  1.01$\pm$  0.06&  1.75$\pm$  0.04&  0.24$\pm$  0.02&  2.05$\pm$  0.08&  1.14$\pm$  0.05&  0.56&  0.9& 44\\
V592 Per& A   0.716&     6800$\pm$    40&  6020$\pm$    40&  2.25$\pm$  0.03&  1.47$\pm$  0.02&  9.58$\pm$  0.21&  2.50$\pm$  0.09&  1.74$\pm$  0.06&  0.68$\pm$  0.04&  2.03$\pm$  0.15&  1.29$\pm$  0.10&  0.63&  0.9&44 \\
XZ   Leo& A   0.488&     7240$\pm$    13&  6946$\pm$    13&  1.69$\pm$  0.02&  1.00$\pm$  0.01&  6.93$\pm$  0.04&  2.07$\pm$  0.02&  1.74$\pm$  0.05&  0.59$\pm$  0.03&  2.03$\pm$  0.10&  1.26$\pm$  0.06&  0.62&  0.9& 8\\
DK      Cyg& A   0.471&     7500 &  6700 &  1.71 &  0.99 &  8.27 &  1.76 &  1.74 &  0.53 &  2.01 &  1.24 &  0.62&  1.0&37 \\
NN   Vir& A   0.481&     6900$\pm$    15&  6925$\pm$    15&  1.72$\pm$  0.01&  1.25$\pm$  0.01&  5.91$\pm$  0.06&  3.16$\pm$  0.06&  1.73$\pm$  0.02&  0.85$\pm$  0.02&  1.95$\pm$  0.08&  1.36$\pm$  0.06&  0.70&  1.1& 6\\
$\eta$  CrA& A   0.591&     7100 &  6640 &  2.20$\pm$  0.03&  0.85$\pm$  0.01& 10.27 &  1.34 &  1.72$\pm$  0.04&  0.22$\pm$  0.02&  2.18$\pm$  0.06&  1.06$\pm$  0.03&  0.49&  0.8& 9,37\\
RZ      Tau& A   0.416&     7300$\pm$     8&  7194$\pm$     8&  1.56$\pm$  0.07&  1.04$\pm$  0.05&  6.19$\pm$  0.61&  2.60$\pm$  0.27&  1.70$\pm$  0.16&  0.64$\pm$  0.06&  2.10$\pm$  0.25&  1.21$\pm$  0.14&  0.58&  0.9& 40\\
V401    Cyg& A   0.583&     6700 &  6650 &  1.98 &  1.19 &  7.08 &  2.49 &  1.68 &  0.49 &  2.24 &  1.09 &  0.49&  0.7& 37\\
AQ   Psc& A   0.476&     6100$\pm$    15&  6124$\pm$    15&  1.75$\pm$  0.01&  0.89$\pm$  0.01&  3.76$\pm$  0.02&  0.98$\pm$  0.01&  1.68$\pm$  0.03&  0.39$\pm$  0.02&  1.89$\pm$  0.07&  1.18$\pm$  0.04&  0.62&  1.2& 8\\
AH      Aur& A   0.494&     6215$\pm$     9&  6141$\pm$     9&  1.85$\pm$  0.02&  0.89$\pm$  0.01&  4.59 &  1.01$\pm$  0.03&  1.68$\pm$  0.05&  0.28$\pm$  0.01&  2.01$\pm$  0.04&  1.10$\pm$  0.02&  0.55&  1.0& 35,37\\
V839    Oph& A   0.409&     6650 &  6554 &  1.48 &  0.88 &  3.84 &  1.28 &  1.64 &  0.50 &  1.88 &  1.18 &  0.62&  1.2& 37\\
FP   Boo& A   0.640&     6980$\pm$    14&  6456$\pm$    14&  2.31$\pm$  0.03&  0.77$\pm$  0.01& 11.2$\pm$  0.10&  0.92$\pm$  0.01&  1.61$\pm$  0.05&  0.15$\pm$  0.02&  2.09$\pm$  0.06&  0.96$\pm$  0.03&  0.46&  0.9& 8\\
V1073   Cyg& A   0.786&     6700$\pm$   100&  6590$\pm$   100&  2.24$\pm$  0.02&  1.33$\pm$  0.02&  8.91$\pm$  0.62&  2.88$\pm$  0.02&  1.60$\pm$  0.02&  0.51$\pm$  0.01&  2.29$\pm$  0.03&  1.00$\pm$  0.01&  0.44&  0.7&1 \\
EX   Leo& A   0.409&     6340$\pm$    14&  6110$\pm$    14&  1.56$\pm$  0.01&  0.73$\pm$  0.01&  3.47$\pm$  0.09&  0.66$\pm$  0.03&  1.57$\pm$  0.03&  0.31$\pm$  0.02&  1.80$\pm$  0.08&  1.07$\pm$  0.05&  0.59&  1.4& 45\\
AH      Cnc& A   0.360&     6300$\pm$    90&  6275$\pm$    90&  1.40$\pm$  0.09&  0.68$\pm$  0.05&  2.78$\pm$  0.50&  0.64$\pm$  0.11&  1.47$\pm$  0.15&  0.25$\pm$  0.03&  1.85$\pm$  0.16&  0.93$\pm$  0.08&  0.50&  1.3& 39\\
AP      Leo& A   0.430&     6150$\pm$    25&  6250$\pm$    25&  1.48$\pm$  0.05&  0.82$\pm$  0.03&  2.80 &  0.91$\pm$  0.08&  1.46$\pm$  0.04&  0.43$\pm$  0.02&  1.80$\pm$  0.11&  1.00$\pm$  0.06&  0.55&  1.4&21,37\\
FG      Hya& A   0.328&     6200$\pm$    20&  6519$\pm$    20&  1.44$\pm$  0.01&  0.52$\pm$  0.01&  2.70$\pm$  0.04&  0.42$\pm$  0.01&  1.45$\pm$  0.03&  0.16$\pm$  0.01&  1.79$\pm$  0.04&  0.90$\pm$  0.02&  0.50&  1.4&45 \\
UX   Eri& A   0.445&     6100$\pm$    28&  6340$\pm$    28&  1.47$\pm$  0.01&  0.91$\pm$  0.01&  2.64$\pm$  0.04&  1.17$\pm$  0.07&  1.43$\pm$  0.03&  0.53$\pm$  0.02&  1.80$\pm$  0.10&  1.00$\pm$  0.05&  0.56&  1.4& 7\\
YY      CrB& A   0.377&     6135$\pm$    10&  6142$\pm$    10&  1.43$\pm$  0.01&  0.80$\pm$  0.01&  2.59 &  0.82$\pm$  0.03&  1.43$\pm$  0.03&  0.35$\pm$  0.01&  1.85$\pm$  0.05&  0.93$\pm$  0.02&  0.50&  1.3& 28,37\\
CK      Boo& A   0.355&     6200 &  6291 &  1.45 &  0.58 &  2.78 &  0.47 &  1.42 &  0.15 &  1.84 &  0.85 &  0.46&  1.3& 37\\
V566    Oph& A   0.410&     7000$\pm$    40&  6881$\pm$    40&  1.47$\pm$  0.01&  0.79$\pm$  0.01&  4.65$\pm$  0.04&  1.25$\pm$  0.01&  1.40$\pm$  0.03&  0.33$\pm$  0.01&  2.06$\pm$  0.03&  0.82$\pm$  0.01&  0.40&  0.9& 24\\
TYC1174$^a$ & A   0.389&     6500$\pm$    36&  6357$\pm$    36&  1.45$\pm$  0.02&  0.71$\pm$  0.01&  3.31$\pm$  0.42&  0.74$\pm$  0.02&  1.38$\pm$  0.01&  0.26$\pm$  0.01&  1.90$\pm$  0.04&  0.83$\pm$  0.02&  0.44&  1.2& 12\\
CN    Hyi& A   0.456&     6500$\pm$   200&  6360$\pm$   260&  1.60$\pm$  0.02&  0.77$\pm$  0.03&  4.10$\pm$  0.60&  0.87$\pm$  0.22&  1.37$\pm$  0.06&  0.25$\pm$  0.02&  1.98$\pm$  0.16&  0.79$\pm$  0.07&  0.40&  1.0& 25\\
GR      Vir& A   0.347&     6300$\pm$    10&  6163$\pm$     9&  1.42$\pm$  0.07&  0.61$\pm$  0.04&  2.87$\pm$  0.28&  0.48$\pm$  0.06&  1.37$\pm$  0.16&  0.17$\pm$  0.06&  1.83$\pm$  0.23&  0.81$\pm$  0.10&  0.45&  1.3& 29\\
DZ   Psc& A   0.366&     6210$\pm$    12&  6187$\pm$    12&  1.47$\pm$  0.02&  0.62$\pm$  0.01&  2.84$\pm$  0.08&  0.49$\pm$  0.02&  1.35$\pm$  0.06&  0.18$\pm$  0.02&  1.83$\pm$  0.08&  0.80$\pm$  0.03&  0.44&  1.3& 7\\
RT      LMi& A   0.375&     6200$\pm$    58&  6350$\pm$    58&  1.29$\pm$  0.02&  0.80$\pm$  0.02&  2.18$\pm$  0.08&  0.93$\pm$  0.06&  1.31$\pm$  0.05&  0.48$\pm$  0.03&  1.75$\pm$  0.13&  0.88$\pm$  0.07&  0.51&  1.5& 45\\
HN   Uma& A   0.383&     6100$\pm$    77&  6082$\pm$    77&  1.44$\pm$  0.01&  0.58$\pm$  0.01&  2.55$\pm$  0.03&  0.41$\pm$  0.03&  1.28$\pm$  0.06&  0.18$\pm$  0.01&  1.76$\pm$  0.06&  0.75$\pm$  0.02&  0.43&  1.5& 44\\
V410 Aur& A   0.366&     5890$\pm$    22&  5983$\pm$    22&  1.44$\pm$  0.03&  0.59$\pm$  0.01&  2.23$\pm$  0.05&  0.39$\pm$  0.02&  1.27$\pm$  0.06&  0.17$\pm$  0.03&  1.75$\pm$  0.11&  0.74$\pm$  0.05&  0.42&  1.5& 8\\
SX      Crv& A   0.317&     6340$\pm$    32&  6160$\pm$    32&  1.31$\pm$  0.02&  0.44$\pm$  0.01&  2.50 &  0.25$\pm$  0.02&  1.25$\pm$  0.04&  0.10$\pm$  0.01&  1.68$\pm$  0.05&  0.72$\pm$  0.02&  0.43&  1.7& 37,43\\
EQ      Tau& A   0.341&     5860$\pm$    10&  5851$\pm$    10&  1.14$\pm$  0.01&  0.79$\pm$  0.01&  1.37 &  0.65$\pm$  0.02&  1.22$\pm$  0.03&  0.54$\pm$  0.02&  1.47$\pm$  0.09&  0.91$\pm$  0.06&  0.62&  2.6& 28,37\\
Y       Sex& A   0.420&     6210 &  6093 &  1.50$\pm$  0.02&  0.75$\pm$  0.01&  3.00$\pm$  0.44&  0.69$\pm$  0.10&  1.21$\pm$  0.15&  0.22$\pm$  0.03&  1.91$\pm$  0.15&  0.64$\pm$  0.05&  0.33&  1.1& 41\\
V2357 Oph& A   0.416&     5640$\pm$    25&  5780$\pm$    25&  1.39$\pm$  0.02&  0.69$\pm$  0.01&  1.78$\pm$  0.03&  0.47$\pm$  0.01&  1.19$\pm$  0.01&  0.29$\pm$  0.01&  1.68$\pm$  0.04&  0.72$\pm$  0.02&  0.43&  1.7& 8\\
V404  Peg& A   0.419&     6340 &  6154 &  1.35 &  0.71 &  2.62 &  0.65 &  1.18 &  0.29 &  1.82 &  0.67 &  0.37&  1.3& 11\\
QX      And& A   0.412&     6500$\pm$   150&  6421$\pm$   155&  1.40$\pm$  0.01&  0.70$\pm$  0.01&  3.12 &  0.74$\pm$  0.09&  1.18$\pm$  0.17&  0.24$\pm$  0.04&  1.92$\pm$  0.17&  0.61$\pm$  0.05&  0.32&  1.1& 31\\
VZ      Lib& A   0.358&     5770$\pm$    21&  5980$\pm$    21&  1.17$\pm$  0.05&  0.72$\pm$  0.03&  1.36$\pm$  0.19&  0.59$\pm$  0.05&  1.06$\pm$  0.06&  0.35$\pm$  0.03&  1.71$\pm$  0.13&  0.60$\pm$  0.05&  0.35&  1.6&33 \\
OO      Aql& A   0.507&     5700$\pm$   300&  5680$\pm$   300&  1.38$\pm$  0.02&  1.28$\pm$  0.02&  1.82$\pm$  0.38&  1.55$\pm$  0.32&  1.05$\pm$  0.02&  0.88$\pm$  0.02&  1.30$\pm$  0.33&  0.91$\pm$  0.23&  0.70&  3.9& 18\\
V508    Oph& A   0.345&     6000 &  5830 &  1.06 &  0.80 &  1.30 &  0.66 &  1.01 &  0.52 &  1.51 &  0.68 &  0.45&  2.4& 37\\
DX      Tuc& A   0.377&     6250$\pm$    37&  6182$\pm$    37&  1.20$\pm$  0.04&  0.71$\pm$  0.02&  1.97$\pm$  0.25&  0.66$\pm$  0.04&  1.00$\pm$  0.03&  0.30$\pm$  0.01&  1.81$\pm$  0.06&  0.49$\pm$  0.02&  0.27&  1.3& 33\\
TV      Mus& A   0.446&     5980$\pm$    32&  6088$\pm$    32&  1.47$\pm$  0.17&  0.67$\pm$  0.08&  2.48$\pm$  0.41&  0.55$\pm$  0.10&  0.94$\pm$  0.14&  0.13$\pm$  0.02&  1.91$\pm$  0.12&  0.34$\pm$  0.02&  0.18&  1.1& 15\\
XY      Boo& A   0.371&     7200 &  7102 &  1.28$\pm$  0.05&  0.55$\pm$  0.02&  3.94 &  0.69 &  0.93$\pm$  0.34&  0.15$\pm$  0.05&  1.98$\pm$  0.15&  0.32$\pm$  0.02&  0.16&  1.0& 2,37\\
TZ      Boo& A   0.298&     5888$\pm$   271&  5754$\pm$   265&  0.96$\pm$  0.13&  0.40$\pm$  0.05&  0.98$\pm$  0.16&  0.16$\pm$  0.03&  0.72$\pm$  0.05&  0.15$\pm$  0.04&  1.47$\pm$  0.19&  0.28$\pm$  0.04&  0.19&  2.6& 14\\

DN      Cam& W   0.498&     6530$\pm$    23&  6700$\pm$    23&  1.78$\pm$  0.02&  1.22$\pm$  0.01&  5.06$\pm$  0.13&  2.67$\pm$  0.06&  1.85$\pm$  0.02&  0.82$\pm$  0.02&  1.88$\pm$  0.09&  1.50$\pm$  0.07&  0.80&  1.2& 3\\
V728    Her& W   0.471&     6622 &  6787 &  1.81 &  0.92 &  5.65 &  1.59 &  1.65 &  0.30 &  2.19 &  1.01 &  0.46&  0.8& 37\\
V402    Aur& W   0.604&     6700$\pm$    31&  6775$\pm$    31&  1.98$\pm$  0.02&  0.92$\pm$  0.01&  7.05 &  1.75$\pm$  0.07&  1.64$\pm$  0.05&  0.33$\pm$  0.02&  2.21$\pm$  0.08&  1.01$\pm$  0.03&  0.46&  0.7& 37,43\\
EF      Boo& W   0.430&     6338 &  6450 &  1.46 &  1.09 &  3.08 &  1.84 &  1.61 &  0.82 &  1.60 &  1.35 &  0.84&  2.0& 37\\
ET    Leo& W  0.347&     5112$\pm$    28&  5500$\pm$    28&  1.36$\pm$  0.01&  0.84$\pm$  0.01&  1.12$\pm$  0.02&  0.56$\pm$  0.01&  1.59$\pm$  0.02&  0.54$\pm$  0.01&  1.39$\pm$  0.05&  1.30$\pm$  0.05&  0.94&  3.1& 8\\
AA      UMa& W   0.468&     5929 &  5965 &  1.47 &  1.11 &  2.39 &  1.40 &  1.56 &  0.85 &  1.29 &  1.41 &  1.09&  4.0& 37\\
YY      Eri& W   0.321&     5370$\pm$   250&  5623$\pm$   260&  1.20$\pm$  0.06&  0.77$\pm$  0.04&  1.07$\pm$  0.17&  0.52$\pm$  0.08&  1.55$\pm$  0.14&  0.62$\pm$  0.07&  1.16$\pm$  0.45&  1.37$\pm$  0.53&  1.18&  5.7& 14\\
ER      Ori& W   0.423&     6200 &  6314 &  1.39 &  1.14 &  2.56 &  1.85 &  1.53 &  0.98 &  1.05 &  1.51 &  1.44&  8.1& 10,37\\
V502    Oph& W   0.453&     5888$\pm$   271&  6166$\pm$   284&  1.51$\pm$  0.14&  0.93$\pm$  0.06&  2.45$\pm$  0.40&  1.12$\pm$  0.18&  1.51$\pm$  0.28&  0.56$\pm$  0.08&  1.73$\pm$  0.36&  1.12$\pm$  0.23&  0.64&  1.5& 14\\
V870    Ara& W   0.400&     5860$\pm$    35&  6210$\pm$    35&  1.67$\pm$  0.01&  0.61$\pm$  0.01&  2.96$\pm$  0.30&  0.50$\pm$  0.01&  1.50$\pm$  0.01&  0.12$\pm$  0.01&  1.89$\pm$  0.04&  0.91$\pm$  0.02&  0.48&  1.2& 33\\
BB    Peg& W  0.362&     5780$\pm$     7&  6100$\pm$     7&  1.28$\pm$  0.01&  0.81$\pm$  0.01&  1.61$\pm$  0.03&  0.81$\pm$  0.01&  1.42$\pm$  0.02&  0.55$\pm$  0.01&  1.57$\pm$  0.04&  1.08$\pm$  0.03&  0.68&  2.1&44 \\
VY    Sex& W  0.443&     5756$\pm$    15&  5960$\pm$    15&  1.50$\pm$  0.01&  0.86$\pm$  0.01&  2.17$\pm$  0.02&  0.83$\pm$  0.01&  1.42$\pm$  0.02&  0.45$\pm$  0.01&  1.73$\pm$  0.04&  0.99$\pm$  0.02&  0.57&  1.5& 8\\
V417    Aql& W   0.370&     6030 &  6256 &  1.31 &  0.84 &  2.02 &  0.96 &  1.40 &  0.50 &  1.74 &  0.98 &  0.57&  1.5& 37\\
AE      Phe& W   0.362&     5880$\pm$   270&  6160$\pm$   280&  1.29$\pm$  0.03&  0.81$\pm$  0.02&  1.74$\pm$  0.28&  0.83$\pm$  0.13&  1.38$\pm$  0.06&  0.63$\pm$  0.02&  1.44$\pm$  0.18&  1.11$\pm$  0.14&  0.77&  2.8& 14\\
EZ      Hya& W   0.450&     5721 &  6100 &  1.54 &  0.85 &  2.28 &  0.90 &  1.37 &  0.35 &  1.89 &  0.85 &  0.45&  1.2& 37\\
UV      Lyn& W   0.415&     6045$\pm$    12&  6262$\pm$    12&  1.39$\pm$  0.01&  0.89$\pm$  0.01&  2.30 &  1.09$\pm$  0.03&  1.36$\pm$  0.02&  0.50$\pm$  0.01&  1.80$\pm$  0.05&  0.92$\pm$  0.02&  0.51&  1.4& 35,37\\
AH      Vir& W   0.407&     5300 &  5671 &  1.41 &  0.84 &  1.40 &  0.65 &  1.36 &  0.41 &  1.67 &  0.94 &  0.56&  1.7& 37\\
V842    Her& W   0.419&     6000 &  6280 &  1.46 &  0.81 &  2.47 &  0.92 &  1.36 &  0.35 &  1.90 &  0.84 &  0.44&  1.2& 37 \\
W       UMa& W   0.334&     6026$\pm$   278&  6310$\pm$   291&  1.17$\pm$  0.03&  0.85$\pm$  0.02&  1.62$\pm$  0.26&  1.02$\pm$  0.17&  1.35$\pm$  0.09&  0.69$\pm$  0.05&  1.45$\pm$  0.32&  1.09$\pm$  0.24&  0.75&  2.7& 14\\
QW      Gem& W   0.358&     5890$\pm$    10&  6100$\pm$    10&  1.26$\pm$  0.03&  0.75$\pm$  0.02&  1.71 &  0.69$\pm$  0.04&  1.31$\pm$  0.04&  0.44$\pm$  0.01&  1.66$\pm$  0.06&  0.90$\pm$  0.03&  0.54&  1.8& 21,37\\
SS      Ari& W   0.406&     5860 &  6123 &  1.36 &  0.80 &  1.96 &  0.81 &  1.31 &  0.40 &  1.78 &  0.84 &  0.47&  1.4& 37\\
V752    Cen& W   0.370&     5955 &  6221 &  1.28 &  0.75 &  1.85 &  0.76 &  1.30 &  0.40 &  1.76 &  0.84 &  0.48&  1.5& 4,37\\
AM  Leo& W   0.366&     6100$\pm$     5&  6221$\pm$     5&  1.23$\pm$  0.01&  0.85$\pm$  0.01&  1.84$\pm$  0.02&  0.95$\pm$  0.01&  1.29$\pm$  0.01&  0.59$\pm$  0.01&  1.59$\pm$  0.04&  0.95$\pm$  0.03&  0.60&  2.0& 45 \\
V781    Tau& W   0.408&     5885$\pm$    50&  6150$\pm$    50&  1.21$\pm$  0.01&  0.85$\pm$  0.01&  1.58$\pm$  0.04&  0.93$\pm$  0.04&  1.29$\pm$  0.07&  0.57$\pm$  0.03&  1.62$\pm$  0.13&  0.94$\pm$  0.08&  0.58&  1.9&18,38 \\
V1191 Cyg& W  0.313&     6500 &  6610 &  1.31 &  0.52 &  2.71 &  0.46 &  1.29$\pm$  0.08&  0.13$\pm$  0.01&  1.85$\pm$  0.03&  0.71$\pm$  0.01&  0.38&  1.3& 34\\
RZ      Com& W   0.339&     6166$\pm$   284&  6457$\pm$   298&  1.12$\pm$  0.03&  0.78$\pm$  0.02&  1.62$\pm$  0.26&  0.93$\pm$  0.15&  1.23$\pm$  0.09&  0.55$\pm$  0.04&  1.65$\pm$  0.23&  0.86$\pm$  0.12&  0.52&  1.8& 14 \\
GM    Dra& W  0.339&     6306$\pm$    58&  6450$\pm$    58&  1.25$\pm$  0.01&  0.61$\pm$  0.01&  2.19$\pm$  0.17&  0.56$\pm$  0.01&  1.21$\pm$  0.04&  0.22$\pm$  0.02&  1.83$\pm$  0.07&  0.67$\pm$  0.02&  0.36&  1.3& 7\\
FU      Dra& W   0.307&     5800$\pm$     8&  6133$\pm$     8&  1.12$\pm$  0.01&  0.61$\pm$  0.01&  1.27 &  0.46$\pm$  0.02&  1.17$\pm$  0.04&  0.29$\pm$  0.01&  1.68$\pm$  0.05&  0.70$\pm$  0.02&  0.42&  1.7& 35,37\\
U       Peg& W   0.375&     5860$\pm$    10&  5841$\pm$    10&  1.22$\pm$  0.01&  0.74$\pm$  0.01&  1.58 &  0.58$\pm$  0.02&  1.15$\pm$  0.01&  0.38$\pm$  0.01&  1.66$\pm$  0.05&  0.72$\pm$  0.02&  0.43&  1.8& 28,37\\
GZ      And& W   0.305&     5810$\pm$    25&  6200$\pm$    25&  1.01$\pm$  0.01&  0.74$\pm$  0.01&  1.03 &  0.73$\pm$  0.03&  1.12$\pm$  0.02&  0.59$\pm$  0.02&  1.44$\pm$  0.10&  0.83$\pm$  0.06&  0.58&  2.8&3,37 \\
AO      Cam& W   0.330&     5590$\pm$    33&  5900$\pm$    33&  1.09$\pm$  0.01&  0.73$\pm$  0.01&  1.03$\pm$  0.07&  0.58$\pm$  0.01&  1.12$\pm$  0.01&  0.49$\pm$  0.01&  1.50$\pm$  0.05&  0.78$\pm$  0.02&  0.52&  2.5& 3,7\\
TW      Cet& W   0.317&     5450 &  5630 &  0.99 &  0.76 &  0.77 &  0.52 &  1.06 &  0.61 &  1.18 &  0.87 &  0.73&  5.3& 37\\
BV      Dra& W   0.350&     6245$\pm$    30&  6345$\pm$    30&  1.12$\pm$  0.01&  0.76$\pm$  0.01&  1.71 &  0.84$\pm$  0.04&  1.04$\pm$  0.02&  0.43$\pm$  0.01&  1.76$\pm$  0.05&  0.59$\pm$  0.02&  0.34&  1.5& 19,37\\
BX      Peg& W   0.280&     5300$\pm$    10&  5528$\pm$    10&  0.97$\pm$  0.01&  0.62$\pm$  0.01&  0.66$\pm$  0.01&  0.32$\pm$  0.01&  1.02$\pm$  0.12&  0.38$\pm$  0.05&  1.40$\pm$  0.20&  0.68$\pm$  0.10&  0.48&  3.1& 32\\
OU      Ser& W   0.297&     5960$\pm$    42&  6380$\pm$    42&  1.09$\pm$  0.01&  0.51$\pm$  0.01&  1.34 &  0.38$\pm$  0.02&  1.02$\pm$  0.01&  0.18$\pm$  0.01&  1.73$\pm$  0.05&  0.50$\pm$  0.02&  0.29&  1.6& 27,37\\
AB      And& W   0.332&     5495$\pm$   250&  5888$\pm$   270&  1.05$\pm$  0.01&  0.76$\pm$  0.01&  0.87$\pm$  0.20&  0.59$\pm$  0.12&  1.01$\pm$  0.02&  0.49$\pm$  0.01&  1.50$\pm$  0.14&  0.67$\pm$  0.06&  0.45&  2.4&16 \\
SW      Lac& W   0.321&     5347 &  5630 &  1.03 &  0.94 &  0.78 &  0.80 &  0.98 &  0.78 &  0.94 &  0.93 &  0.99& 11.7& 37\\
TY      Boo& W   0.317&     5800$\pm$   200&  6180$\pm$   200&  1.00$\pm$  0.01&  0.69$\pm$  0.01&  1.02$\pm$  0.14&  0.62$\pm$  0.09&  0.93$\pm$  0.02&  0.40$\pm$  0.01&  1.66$\pm$  0.10&  0.51$\pm$  0.03&  0.30&  1.8& 30\\
VW      Cep& W   0.278&     5012$\pm$   231&  5248$\pm$   242&  0.91$\pm$  0.02&  0.62$\pm$  0.01&  0.47$\pm$  0.08&  0.26$\pm$  0.04&  0.93$\pm$  0.02&  0.40$\pm$  0.01&  1.27$\pm$  0.11&  0.64$\pm$  0.05&  0.50&  4.2& 14\\
BW      Dra& W   0.292&     5980$\pm$    10&  6164$\pm$    10&  0.98$\pm$  0.01&  0.55$\pm$  0.01&  1.10 &  0.39$\pm$  0.02&  0.92$\pm$  0.02&  0.26$\pm$  0.01&  1.65$\pm$  0.05&  0.45$\pm$  0.01&  0.27&  1.8& 19,37\\
TX      Cnc& W   0.383&     5888$\pm$   271&  6165$\pm$   284&  1.12$\pm$  0.03&  0.83$\pm$  0.04&  1.35$\pm$  0.22&  0.91$\pm$  0.15&  0.91$\pm$  0.10&  0.50$\pm$  0.06&  1.71$\pm$  0.29&  0.50$\pm$  0.08&  0.29&  1.6& 14\\
V757    Cen& W   0.343&     5900 &  6000 &  0.97 &  0.80 &  1.02 &  0.74 &  0.88 &  0.59 &  1.45 &  0.59 &  0.41&  2.7& 37\\
V829    Her& W   0.358&     5900$\pm$    21&  5380$\pm$    21&  1.07$\pm$  0.01&  0.74$\pm$  0.01&  1.25 &  0.42$\pm$  0.02&  0.86$\pm$  0.02&  0.37$\pm$  0.01&  1.53$\pm$  0.05&  0.47$\pm$  0.02&  0.31&  2.3& 37,43\\
CC      Com& W   0.221&     4200$\pm$   180&  4300$\pm$   180&  0.71$\pm$  0.01&  0.53$\pm$  0.01&  0.14$\pm$  0.01&  0.09$\pm$  0.01&  0.72$\pm$  0.02&  0.38$\pm$  0.01&  0.86$\pm$  0.11&  0.56$\pm$  0.07&  0.65& 15.9&20 \\
XY      Leo& W   0.284&     4361$\pm$     8&  4800$\pm$     8&  0.83$\pm$  0.13&  0.66$\pm$  0.10&  0.22$\pm$  0.08&  0.21$\pm$  0.07&  0.76$\pm$  0.15&  0.46$\pm$  0.06&  1.03$\pm$  0.46&  0.57$\pm$  0.26&  0.55&  8.5&33 \\
V523    Cas& W   0.234&     4410$\pm$     5&  4736$\pm$     5&  0.74$\pm$  0.04&  0.55$\pm$  0.02&  0.18$\pm$  0.02&  0.13$\pm$  0.01&  0.75$\pm$  0.03&  0.38$\pm$  0.02&  1.00$\pm$  0.13&  0.54$\pm$  0.07&  0.55&  9.6&42 \\
BH      Cas& W   0.406&     4790$\pm$   100&  4980$\pm$   100&  1.05$\pm$  0.01&  0.75$\pm$  0.01&  0.52$\pm$  0.01&  0.31$\pm$  0.01&  0.74$\pm$  0.06&  0.35$\pm$  0.03&  1.43$\pm$  0.12&  0.38$\pm$  0.03&  0.26&  2.8& 22,37\\
RW      Dor& W   0.285&     4780 &  5200 &  0.80 &  0.67 &  0.30 &  0.30 &  0.64 &  0.43 &  1.28 &  0.35 &  0.28&  4.1& 37\\
RW      Com& W   0.237&     5120$\pm$    10&  5400$\pm$    10&  0.71$\pm$  0.01&  0.46$\pm$  0.01&  0.31 &  0.16$\pm$  0.01&  0.56$\pm$  0.06&  0.20$\pm$  0.03&  1.41$\pm$  0.12&  0.15$\pm$  0.01&  0.11&  3.0& 23,37\\
\hline
\end{longtable}
%% Any table notes must follow the \end{tabular} command.
%\end{landscape}
%\end{sidewaystable}
%\end{center}
a) TYC 1174-344-1\\
Ref.: 1; \cite{t1}, 2; \cite{t2}, 3; \cite{t3}, 4; \cite{t4}, 5; \cite{t5}, 6; \cite{t6}, 7; \cite{bb1}, 8; \cite{bb2}, 9; \cite{t9}, 10; \cite{t10}, 11; \cite{t11}, 12; \cite{t12}, 13; \cite{t13}, 14; \cite{b3}, 15; \cite{t15}, 16; \cite{t16}, 17; \cite{t17}, 18; \cite{t18}, 19; \cite{t19}, 20; \cite{koseCC},21; \cite{b8}, 22; \cite{t21}, 23; \cite{t22}, 24; \cite{t23}, 25; \cite{t24}, 26; \cite{t25}, 27; \cite{t26}, 28; \cite{t27}, 29; \cite{t28}, 30; \cite{t29}, 31; \cite{t30}, 32; \cite{t31}, 33; \cite{t32}, 34; \cite{t33}, 35; \cite{t34}, 36; \cite{t35}, 37; \cite{b18}, 38; \cite{t37}, 39; \cite{t38}, 40; \cite{t39}, 41; \cite{t40}, 42; \cite{t41}, 43; \cite{t42}, 44; \cite{b21}, 45; \cite{b20}
\twocolumn

\label{lastpage}

\end{document}